\documentclass[aps,showkeys,showpacs,nosuperscriptaddress,twocolumn]{revtex4}
\usepackage{graphicx}
\usepackage{amsmath}
\usepackage{hyperref}
\usepackage{color}
\usepackage{subfigure}
 
\begin{document}

\title{Separate or perish - the coevolving voter model}

\author{Krzysztof Ku{\l}akowski}
\email{kulakowski@fis.agh.edu.pl}
\affiliation{
Faculty of Physics and Applied Computer Science, AGH University of Science and Technology, al. Mickiewicza 30, PL-30059 Krak\'ow, Poland
}

\author{Maria Stojkow}
\email{stojkoff@gmail.com}
\affiliation{
Faculty of Humanities, AGH University of Science and Technology, al. Mickiewicza 30, PL-30059 Krak\'ow, Poland
}

\author{Dorota \.Zuchowska-Skiba}
\email{zuchowskadorota@gmail.com}
\affiliation{
Faculty of Humanities, AGH University of Science and Technology, al. Mickiewicza 30, PL-30059 Krak\'ow, Poland
}
 
\author{Przemys{\l}aw Gawro\'nski}
\email{gawron@agh.edu.pl}
\affiliation{
Faculty of Physics and Applied Computer Science, AGH University of Science and Technology, al. Mickiewicza 30, PL 30-059 Krak\'ow, Poland
} 
 
\date{\today}
 
\begin{abstract}
Recent generalization of the coevolving voter model (J. Toruniewska et al, PRE 96 (2017) 042306) is further generalized here, including spin-dependent probability of rewiring. Mean field results indicate that either the system splits into two separate networks with different spins, or one of spin orientation goes extinct. In both cases, the density of active links is equal to zero. The results are discussed in terms of homophily in social contacts.
\end{abstract}
 
%\pacs{89.65.-s; 07.05.Tp}
 

\maketitle
  
%% ###########################################################################
\section{Introduction}
\label{S0}
%% ###########################################################################
In the coevolving voter model, an actor placed at a node of a social network either copies the state of his/her neighbor, or rewires the link as to connect another node in the same state as him/herself \cite{vzqz}. In the related Monte Carlo scheme, this decision is taken by randomly selected actors. The related probability of rewiring is $p$, and the probability of copying is $1-p$. In the mean-field scheme, differential equations have been written \cite{vzqz} as to evaluate the process averaged over many decisions. The model is known to be useful for modeling catalytic reactions and formation of public opinion \cite{cfl,frkr}. In its standard version, two states of actors are allowed and the mean degree of a node is uniform \cite{vzqz,cfl,vzqe}. Recently, the model has been generalized to the case where the mean degree of a node depends on its state \cite{toru}. Here we present further generalization of the mean field calculations, where both mean degrees $\mu_a, \mu_b$ and the rewiring probabilities $p_a, p_b$ depend on the node states $a,b$. The derivation is exactly the same as in \cite{toru}, then we do not copy it here; it is only that $p_a, p_b$ are used instead of $p$.\\

Here we are particularly interested in the interpretation of the action of rewiring as a demonstration of homophily. According to \cite{mimc}, 'homophily is the principle, that a contact between similar people occurs at a higher rate that among dissimilar people'. In Section \ref{S1} we argue that both rewiring and state copying have their counterparts in social reality. In Section \ref{S2} we provide definitions of the model parameters. Two subsequent sections (\ref{S3} and \ref{S4}) are devoted to analytical and numerical results of the model. In the last Section \ref{S5} we discuss the main results.

\section{Sociological background}
\label{S1}

In the process of simulated rewiring, individuals in both groups show a tendency to limit their social contacts to their own groups. This conforms to the Schelling segregation model. Thomas Schelling placed the separation process in the spontaneous activities of individuals who want to be surrounded by people similar to themselves. The effect of these individual choices was the creation of unintended patterns of separation in the social structure \cite{tcs1}. As a result, the segregation between the two groups was the result of relatively mild individual preferences to live among neighbors similar to each other. This process was strengthened when staying with other groups was indifferent or uncomfortable for the individual \cite{wfs}. \\

Scheling's model has been also confirmed in the situation when group members are open to integration \cite{tcs2,ban}. The results of analyzes conducted by Junfu Zhang \cite{jnf1} regarding the spatial segregation process showed that separation occurs even when the majority of respondents declare that they prefer to live in integrated districts. This makes it possible to explain the mechanism of persistence  of segregation in societies that become more and more tolerant \cite{jnf2}. The research carried out by Fossetta and Dietrich
\cite{fsd} show that segregation processes based on individual choices related to being surrounded by people similar to each other is so strong that its occurrence does not depend on structure of the city in which it happens . In addition, they are characterized by high durability, because as shown in \cite{wfs}, changes of individual preferences  in this area are very slow.\\

Separation processes based on spontaneous actions of individuals resulting from their individual preferences regarding being surrounded by similar persons constitute a permanent and important factor causing the phenomenon of separation in society \cite{cort,fagi}. They also affect the social network structures built individually by actors, which are based not only on the selection of people who create their nearest surroundings but also on engaging in interactions with other people in accordance with their preferences \cite{fagi}. In this sense, segregation processes based on the Schelling model are the effect of the natural tendency of individuals to homophily, that is, to interact with those who are similar in terms of certain important characteristics, which has consequences for the structure of the network. As a result, personal networks of people are homogeneous in terms of many sociodemographic, behavioral and intrapersonal traits \cite{mimc}. The degree of homophily in social networks influences the level of segregation processes taking place in them. The higher it is, the stronger the tendencies for homophily appear in social networks \cite{bss}. In this sense, homophily contributes to solidifying differences between groups, which deepens the separation between them. In addition, Golub and Jackson \cite{gj} have shown that it slows down the speed at which society reaches consensus.\\

Lazarsfeld and Merton \cite{lzm} distinguished two types of homophily. The first was related to the status of the individual (homophily status) and included basic sociodemographic dimensions that differentiate society, such as: race, ethnicity, gender, age and acquired traits, such as religion, education, occupation or specific behavioral patterns resulting from social status. The second type of homophily was based on values, attitudes and beliefs (value homophily). In this perspective the political orientation is the basis of homophily and can be a factor conducive to separation, which will result in isolation among people sharing the same views.\\

As a result, actors with similar political beliefs will participate in groups corresponding to their preferences. In a situation where an individual changes some personal traits that have so far influenced his membership in a particular group, eg political orientation, the group will be changed \cite{hrst}. This is important for the level of involvement of the actor in his activity in the new group. Similar process can be noticed in the case of conversions, where the people who change affiliation,  in this case religious one,  tend to have clear attitudes in harmony with the new group, often becoming ardent advocates of a new idea or ideology. This is due, among other things, to the fact that people - (e.g. converts, migrants) who lose the frame of reference which so far have provided them with a social position are attracted by new ideas that compensate for those  they lost \cite{elwr}. From the perspective of a convert, previous behavior or belief can be seen as anti-social or felt as a personal threat, resulting in the creation of a social form imagined as a community, but formulated as an organization that excludes immoral and unifies moral, according to new beliefs \cite{snow}.\\

The idea of changing beliefs may be related to the continuum, where radical and total changes are distinguished, based on variations of Nock's distinction \cite{nock} between conversion and adhesion, to indicate the possibility of participating in religious groups and rituals without pursuing a new way of life (in case of adhesion) \cite{maxh}. Therefore, converts tend to be closely integrated with the new group and to identify strongly with it.

\section{The coevolving voter model}
\label{S2}

The medium of our considerations here is a random network, where nodes can be in two states, say $a$ and $b$. It is convenient to call these states as two spin states, up and down. In the coevolutionary voter model \cite{vzqz}, temporal evolution of the network involves two processes, both of them driven by an existence of links between nodes in different states. A randomly selected node either changes its state as to adopt the state of its neighbor, or breaks the link and rewires it to join another node, which is in the same state as itself. It is usually assumed that these actions are performed with probabilities $1-p$ and $p$, respectively. This dynamics ends up in one of two kinds of states: either nodes of different states are completely separated, or there remains some amount of contact. The order parameter of this transition is played by the fraction $\rho$ of links between nodes in different states; they are termed 'active links'. In the former phase, $\rho=0$; in the latter, $\rho$ fluctuates around some value \cite{vzqz}.\\

The coevolving voter model has been generalized recently as to allow for different numbers of neighbors of nodes in different states \cite{toru}. The related mean field model
has been delopped in this way; instead of one differential equation for the density $\rho(t)$ of active links we got three equations for $\rho(t)$, $m(t)$ and $\beta (t)$. There,
the variable $m$ is defined as the order parameter of nodes: $m=(N_a-N_b)/N$, where $N_a$ ($N_b$) is the number of nodes in the state $a$ ($b$), and $N=N_a+N_b$ is the number of all nodes. Further, the variable $\beta$ is defined as the order parameter of links. Namely, we distinguish $M_{aa}$, i.e. the number of links from nodes $a$ directed towards nodes $a$, from the number $M_{ab}$ of links from nodes $a$ directed towards nodes $b$, and so on. Then, we get $M_{aa}-M_{bb}=N\mu\beta$, where $\mu$ is the mean number of neighbors (mean node degree), calculated over the whole network. Also, $M_{ab}=M_{ba}=N\mu \rho/2$. The equations for $m(t)$ and $\beta(t)$ have been found to match in such a way, that a linear combination of these order parameters has been found to remain constant under time evolution \cite{toru}. Also, the subspace defined by the condition 
$\beta=m$ has been found to be invariant: once $\beta$ happened to be equal to $m$, it remains equal forever. All these results have been confirmed by extensive Monte-Carlo simulations \cite{toru}. Yet both mean field and MC method relied on an assumption, that the probability $p$ of rewiring does not depend on the node state.\\

It is precisely this assumption which is relieved here. The motivation of this step is the fact that most applications of the voter model \cite{cfl} and its coevolving versions \cite{vzqz} are phenomenological, as competition between species, opinion formation and catalytic reactions. In such applications, any symmetry either needs justification or remains just a simplifying assumption. This is true in particular in social phenomena, where members of different groups can react in a measurably different way on varying external conditions \cite{curr,barb}. Our aim is to explore the coevolving voter model, where $p_a \ne p_b$: the rewiring probabilities are different for different groups. The method is the mean field approach, as applied in \cite{vzqz,toru}, and the whole approach presented here is a direct continuation of those papers.\\

We note that another approach to the voter model has been aplied recently in \cite{lhdf} from the perspective of voting dynamics and political campaigns. Although the formalism applied there is different from applied here and more focused on game theory, the basic improvements of theoretical descriptions are parallel to our approach. In particular, the dynamics of the network structure reflects the applied strategies of actors at the network nodes, and these strategies depend on the node state.\\

In the next section we provide the obtained equations of motion for $\rho(t)$, $m(t)$ and $\beta (t)$, and a list of symbols which will be useful for interpretation. Section 3 is devoted to the results obtained by the numerical solution of the equations of motion. In the last section we interpret these results in terms of social integration. 

\section{Analytical results}
\label{S3}

We have applied the same scheme of calculation as in Ref. \cite{toru}, with distinguishing the rewiring probabilities $p_a,p_b$ for different nodes as the only modification. The obtained equations of motion are as follows:

\begin{eqnarray}
\frac{d\rho}{dt} & = & \rho \frac{ \Big(-2 + m (2 - p_b) + (1 - \beta) \mu (1 - p_b) + p_b\Big)}{(1 - \beta) } \\ \nonumber
                 & + &  2 \rho^2\frac{ \Big(1 - m - \mu (1 - \beta) \Big) (1 - p_b)}{ (1 - \beta)^2}  \\ \nonumber
    & + & \rho \frac{\Big(-2 - m (2 - p_a) + (1 + \beta) \mu (1 - p_a) + p_a\Big)} {(1 + \beta) }\\ \nonumber
    & + & 2 \rho^2 \frac{ \Big(1 + m - \mu(1 + \beta) \Big) (1 - p_a) )} { (1 + \beta)^2}
\label{rdot}
\end{eqnarray}

\begin{equation}
\frac{dm}{dt}= -\frac{(1+m)(1-p_a) \rho}{1+\beta}  +\frac{(1-m)(1-p_b) \rho}{1-\beta} 
\label{mdot}
\end{equation}

\begin{eqnarray}
\frac{d\beta}{dt} & = & \rho \frac{-(1 + \beta) \Big(1 - m + (1 - \beta) \mu \Big) p_b} {\mu(1 - \beta^2 )}\\ \nonumber
& + &\frac{ (1 - \beta) \Big(1 + m + (1 + \beta )\mu\Big) p_a }  {\mu(1 - \beta^2 )}
\label{bdot}
\end{eqnarray}
Some preliminary conclusions can be drawn from this form of equations, even before they are solved numerically. First, on the contrary to the case $p_a=p_b$ considered 
in \cite{toru}, the subspace $m=\beta$ is not invariant anymore. Instead, putting $m=\beta$ in Eqns. (\ref{mdot},\ref{bdot}) we get

\begin{eqnarray}
\frac{dm}{dt} & = & \rho(p_a-p_b)\nonumber\\
\frac{d\beta}{dt} & = & \frac{\mu+1}{\mu}\rho(p_a-p_b)
\label{meqb}
\end{eqnarray}
This means that $|m-\beta|$  increases in time if initially it is equal to zero. Further, we might like to reproduce the condition of stability of the stationary phase 
$\rho \ne 0$, as it was done in \cite{vzqz}. The reasoning is as follows: let us put $d\rho/dt$ in Eq. (\ref{rdot}) in the form $d\rho/dt=\rho(A+B\rho)$. The boundary of the stability of $\rho = 0$ is where $A=0$. Having substituted $m=\beta$ there, we get a solution $p_a+p_b=2(\mu-2)/(\mu-1)$, what nicely coincides with the results of \cite{vzqz} obtained for $p_a=p_b$. Yet, as shown just above, the condition $m=\beta$ is not justified {\it a priori}. In the next section we will analyse the cases where this condition is fulfilled.\\

All quantities of interest can be expressed by the three variables  $\rho$, $m$ and $\beta$. In particular, let us denote the mean degree of nodes $a,b$ as $\mu_a,\mu_b$. Then we have 

\begin{eqnarray}
\mu_a=\mu \frac{1+\beta}{1+m}\nonumber\\
\mu_b=\mu \frac{1-\beta}{1-m}
\label{muab}
\end{eqnarray}
Again, for $m=\beta$ we get $\mu_a=\mu_b = \mu$; in this limit case there is no topological difference between the nodes $a$ and $b$.\\

\section{Numerical results}
\label{S4}

\begin{figure}[!hptb]
\begin{center}
\includegraphics[width=\columnwidth]{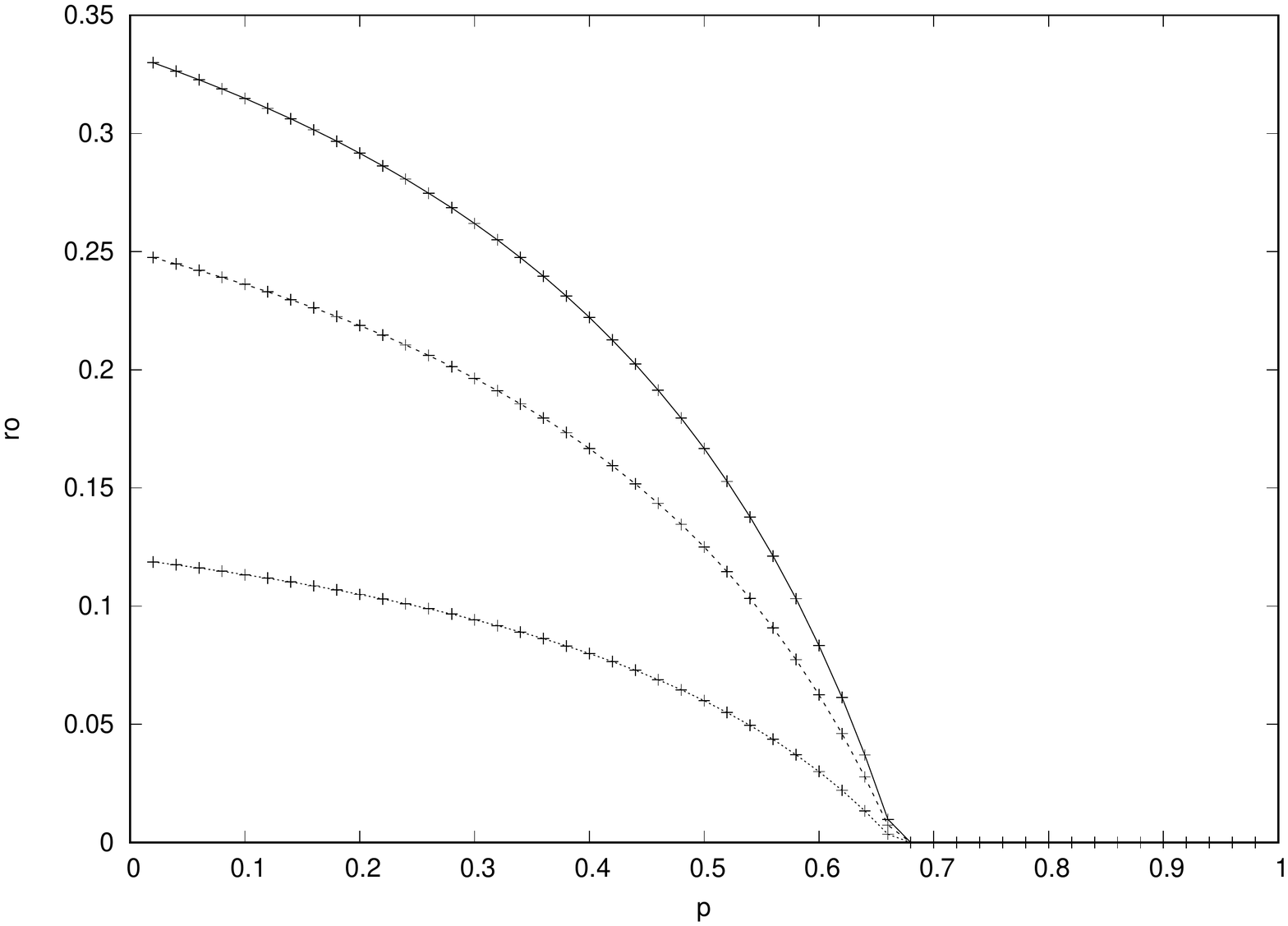}
\caption{Three plots of stationary values of $\rho (p)$ for $p\equiv p_a=p_b$ and $\beta=m$. The plots are made for $m=0$ (upper plot), $m=0.5$ (middle plot), and $m=0.8$ (lower plot). }
\label{f1}
\end{center}
\end{figure}

The mean degree $\mu=4$ is kept for all calculations. As expected, for $p_a=p_b$ and initial $\beta=m$, both $m$ and $\beta$ remain constant and equal to their initial values. The density $\rho$ of active bonds decreases with $p$ and vanishes at $p=2/3$, which agrees with the formula \cite{vzqz} $p_c=(\mu-2)/(\mu-1)$.  The final values of $\rho$ does not change with the initial values of $\rho$, yet it depends on the value of $m$. To demonstrate this, in Fig. (\ref{f1}) we show three plots of $\rho (p)$, for different values of $m$.\\ 

\begin{figure}[!hptb]
\begin{center}
\includegraphics[width=\columnwidth]{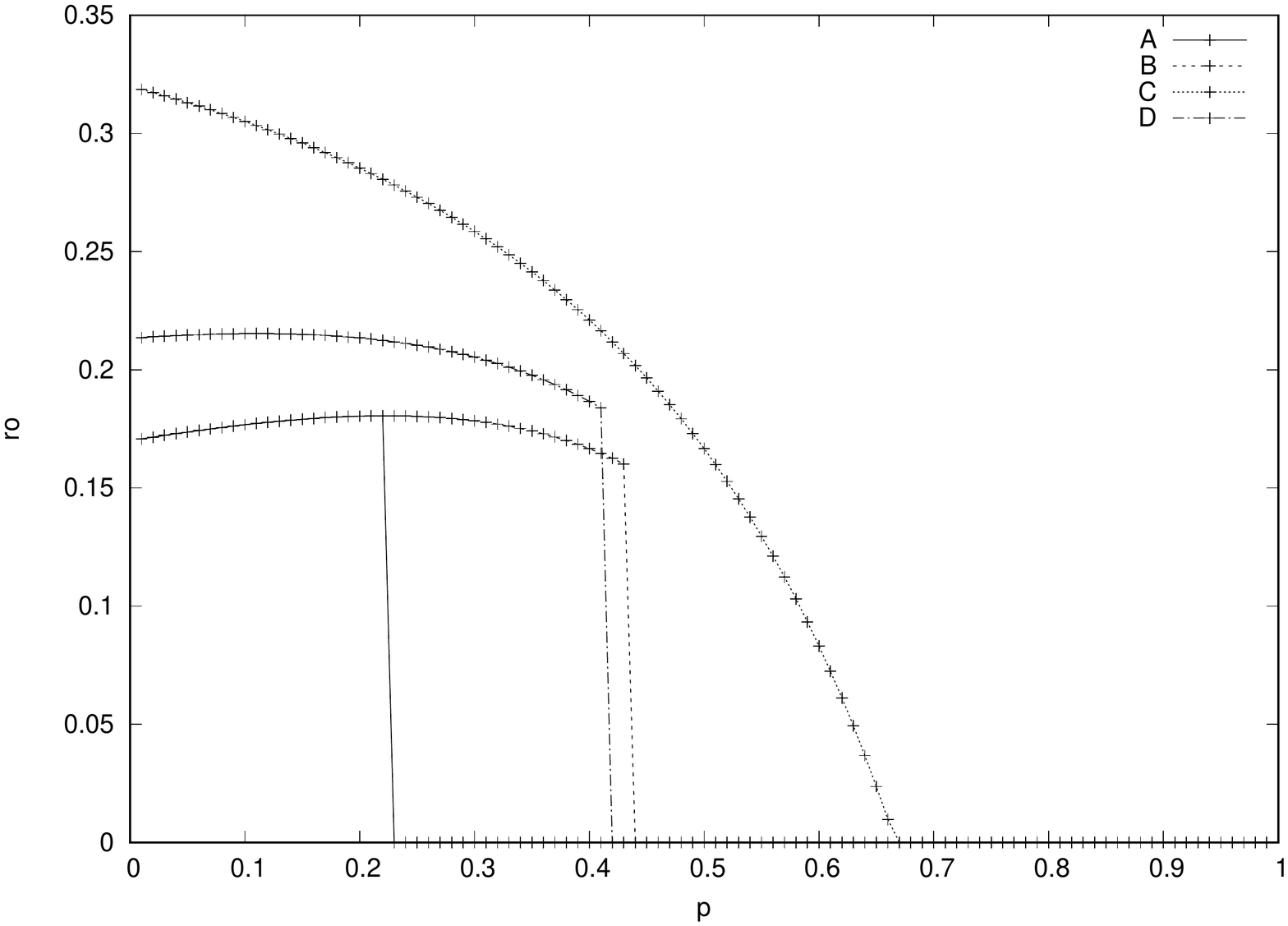}
\caption{The plots of stationary values of $m(p)$ and $\beta (p)$ for $p\equiv p_a=p_b$ and $\beta \ne m$. The plots are made for the initial data as follows: A ($\rho_0 = 0.4$, $m_0 = 0.7$, $\beta _0 = -0.7$) , B ($\rho_0 = 0.8$, $m_0 = 0.7$, $\beta _0 = -0.7$), C ($\rho_0 = 0.8$, $m_0 = 0.7$, $\beta _0 = -0.2$) and D ($\rho_0 = 0.5$, $m_0 = 0.8$, $\beta _0 = -0.6$).}
\label{f2}
\end{center}
\end{figure}

\begin{figure}[!hptb]
\begin{center}
\includegraphics[width=\columnwidth]{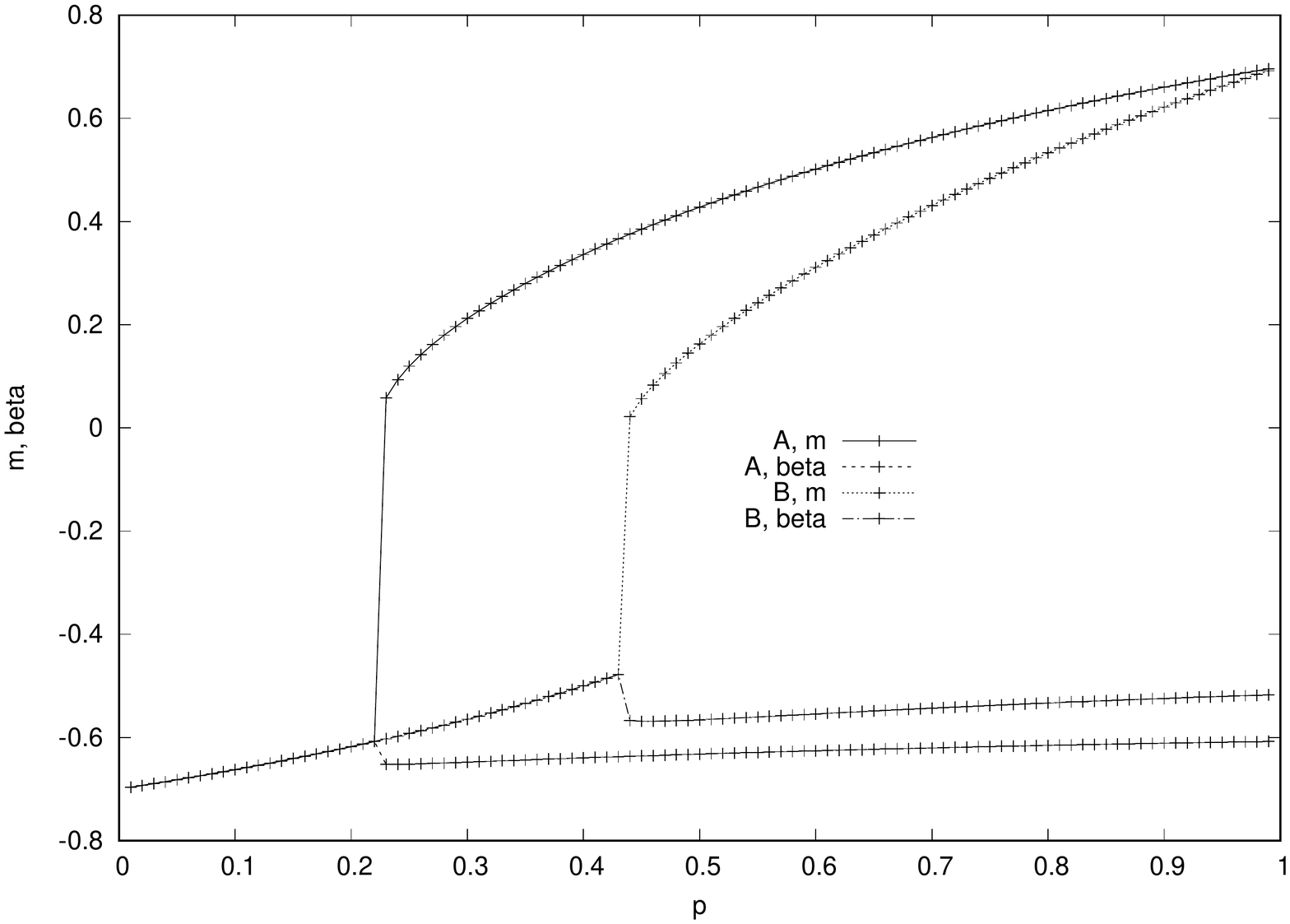}
\includegraphics[width=\columnwidth]{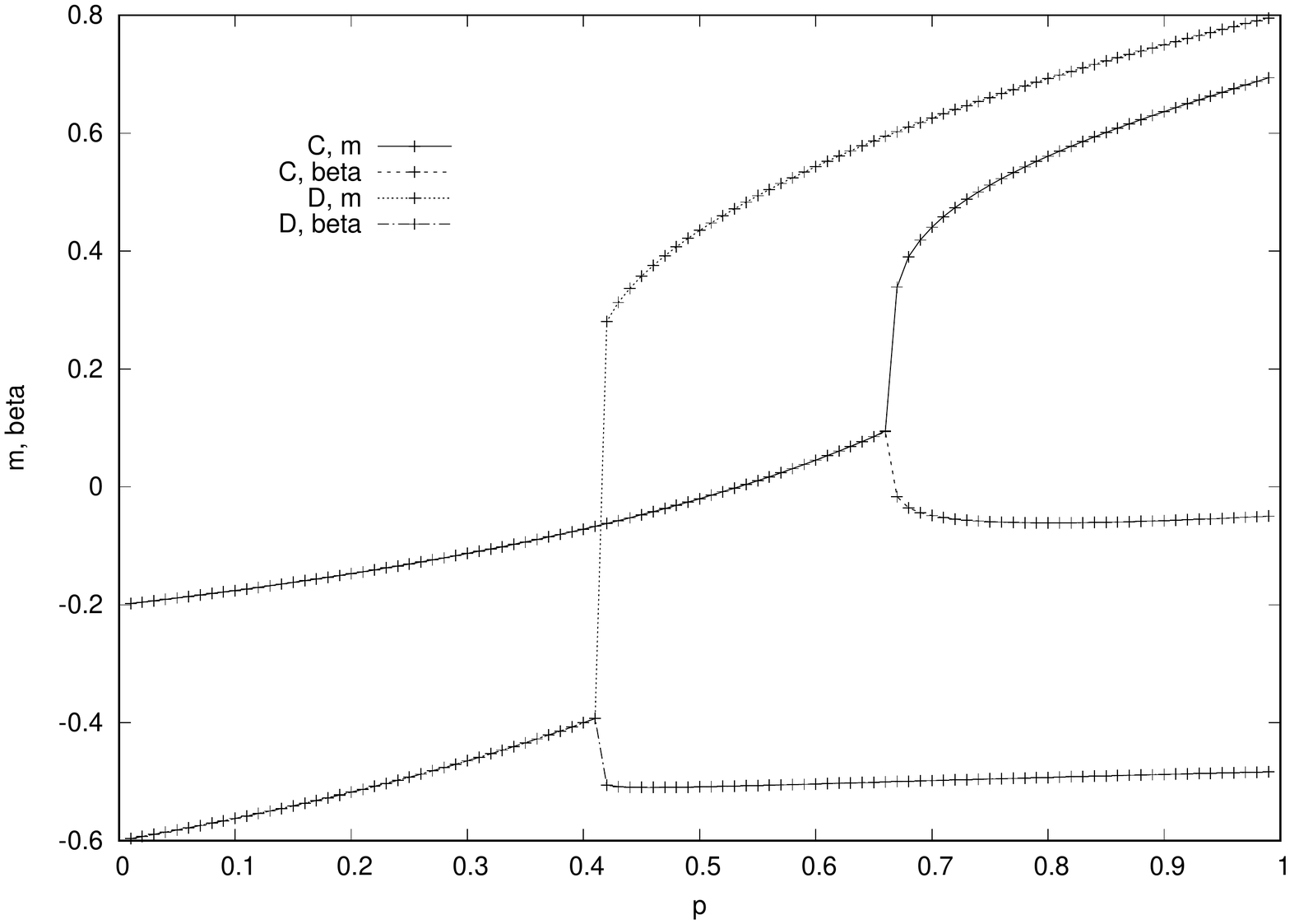}
\caption{The plots of stationary values of $m(p)$ and $\beta (p)$ for $p\equiv p_a=p_b$ and $\beta \ne m$. The plots are made for the initial data as follows: A ($\rho_0 = 0.4$, $m_0 = 0.7$, $\beta _0 = -0.7$) , B ($\rho_0 = 0.8$, $m_0 = 0.7$, $\beta _0 = -0.7$) (upper figure), C ($\rho_0 = 0.8$, $m_0 = 0.7$, $\beta _0 = -0.2$) and D ($\rho_0 = 0.5$, $m_0 = 0.8$, $\beta _0 = -0.6$) (lower figure).}
\label{f3}
\end{center}
\end{figure}

In Fig. (\ref{f2}), the data on stationary $\rho (p)$ are presented for $p\equiv p_a=p_b$, $\beta \ne m$ and various initial conditions. As  we see, a change of the initial value $\rho _0$ of the density of active links leads to a change of the transition probability $p_c$ where $\rho$ vanishes, leaving however the stationary value of $\rho$ unchanged. Further, a change of the initial values $m_0$ and 
$\beta _0$ not only influences the stationary value of $\rho$, but also may change the transition character, continuous to discontinuous or back. This variation is shown in Fig. 
(\ref{f2}). Further, as long as the stationary value of $\rho$ is different from zero, the time evolution drives the system to $m=\beta$, i.e. to the state where the mean degree of all nodes is the same. However, once $\rho$ reaches the zero value, the system evolution is frozen and the values of the order parameters $m$ and $\beta$ remain different. In other words, the mean degrees of nodes depend on the index, $a$ or $b$. This fact is demonstrated in Fig. (\ref{f3}), both for continuous and discontinuous transition. \\

\begin{figure}[!hptb]
\begin{center}
\includegraphics[width=0.75\columnwidth]{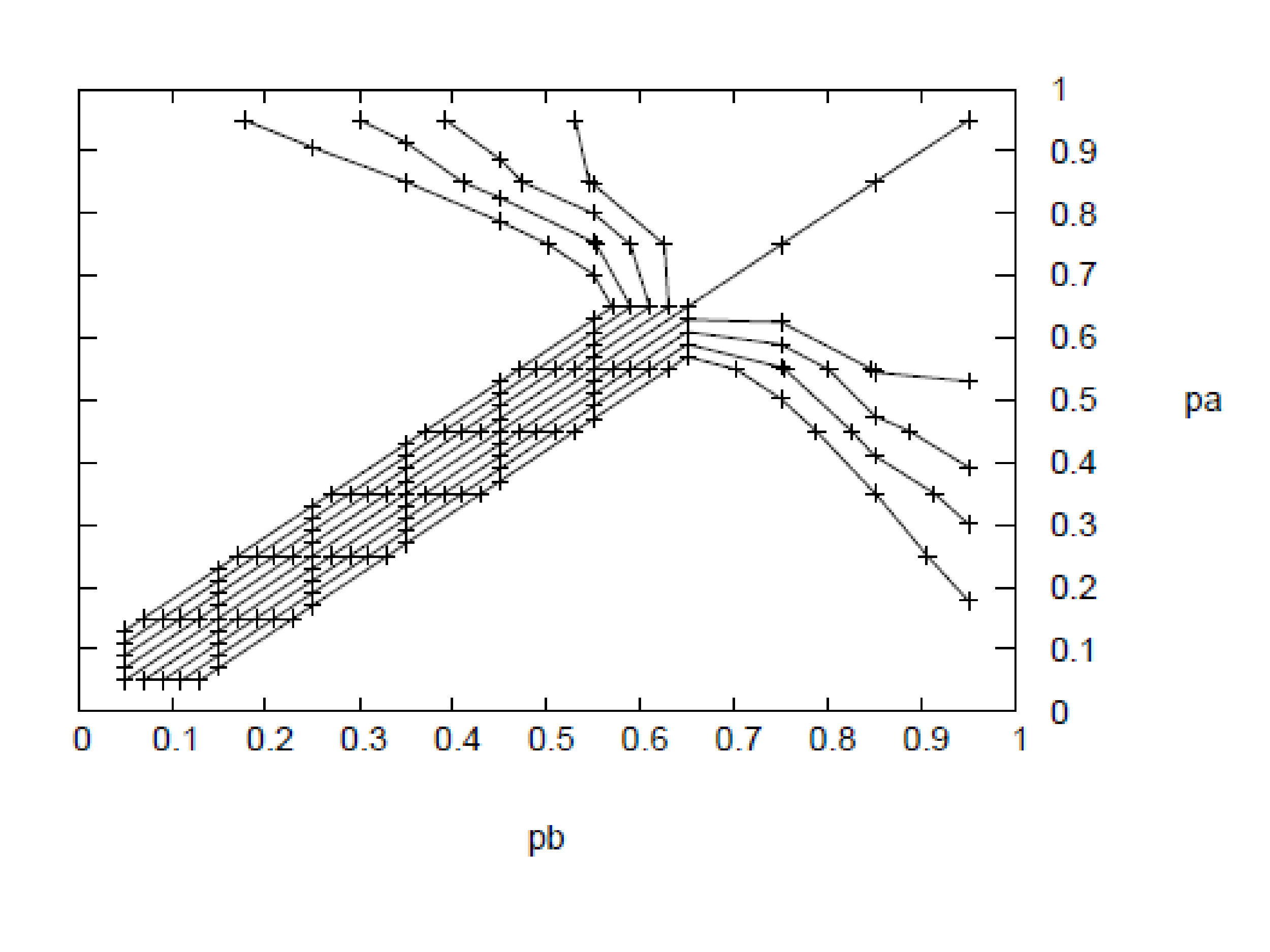}
\includegraphics[width=0.75\columnwidth]{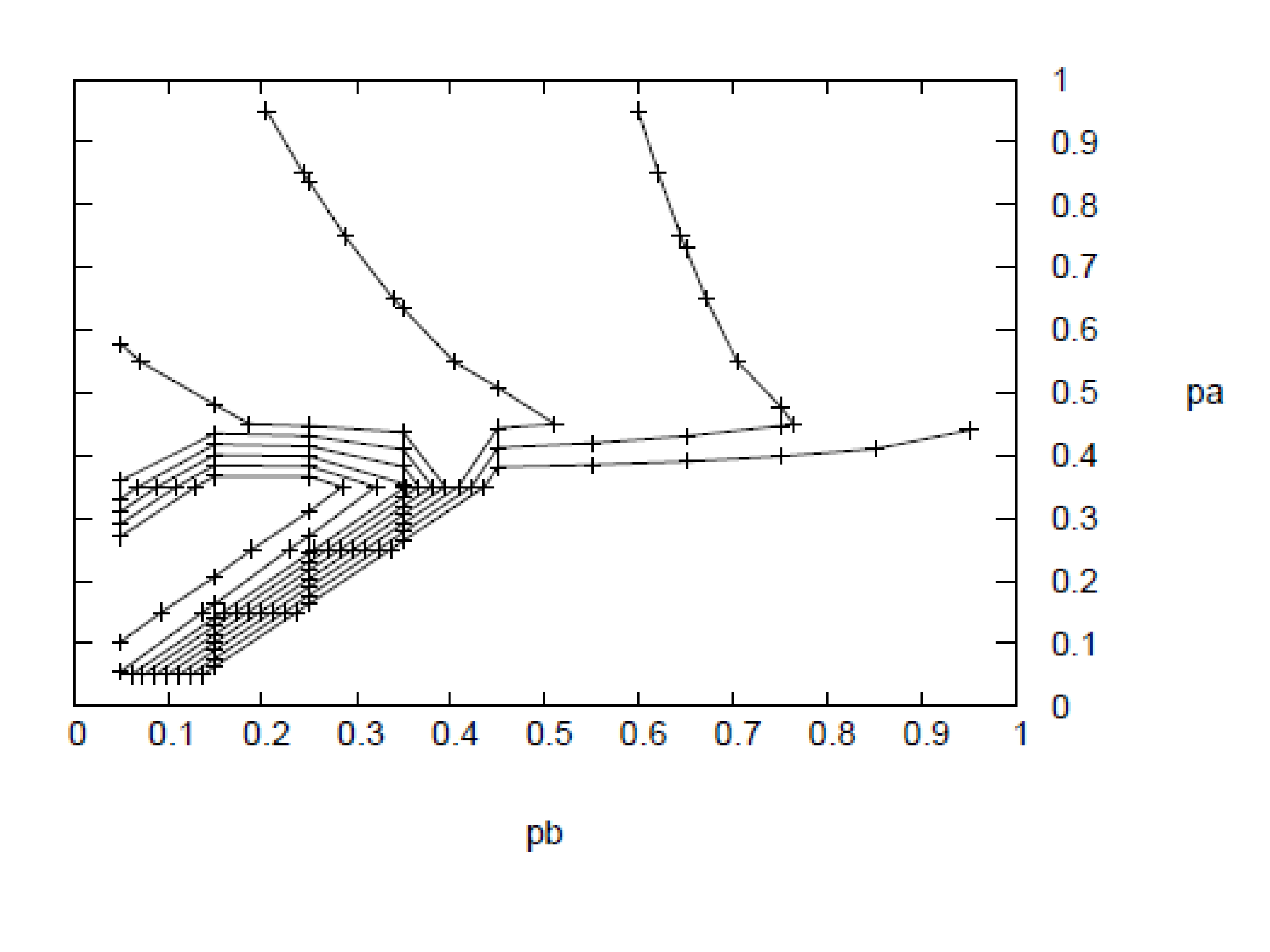}
\includegraphics[width=0.75\columnwidth]{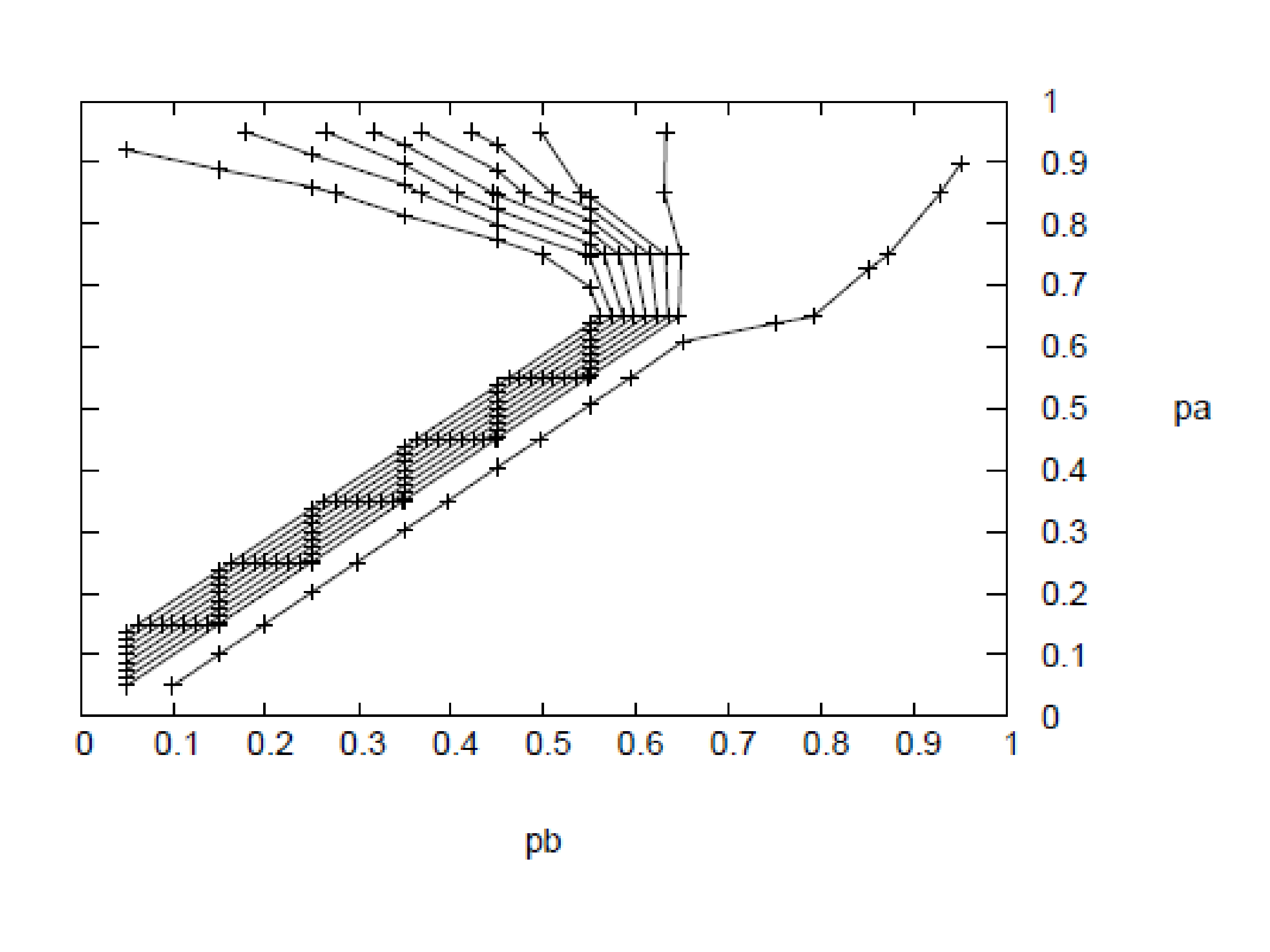}
\caption{Contour maps for the magnetization $m$ vs the position on the plane ($p_a,p_b$) for selected set of initial conditions: $\rho _0=0.5$, $m_0=0.0$, and $\beta _0=0.0$ (upper plot), $\rho _0=0.5$, $m_0=0.8$, and $\beta _0=-0.6$ (middle plot) and $\rho _0=0.5$, $m_0=0.8$, and $\beta _0=0.6$ (lower plot). The lines of constant $m$ are made for 
$m=-0.8,-0.6,-0.4,-0.2,0.0,0.2,0.4,0.6,0.8$.}
\label{f4}
\end{center}
\end{figure}

Our main result is related to the case when $p_a \ne p_b$. Then we get $\rho=0$ in all stationary states: the nodes with spins $a$ are not neighbors of the nodes of spin $b$. This outcome may happen within two scenarios: either $m$ and $\beta$ get frozen or nodes of a given spin disappear. The scenario depends on the initial conditions, i.e. the values of $\rho _0$, $m_0$ and $\beta _0$. In Fig. (\ref{f4}), examples are shown of contour maps of $m$ on the plane ($p_b,p_a$). These maps give an insight on the areas where one of spin orientations is absent or rare. These examples are shown to demonstrate, that an outcome of the time evolution depends on the initial conditions. Yet, the fraction of spins $a$ is always at least 0.1 if $p_a > p_b$,  and the fraction of spins $b$ is always at least 0.1 if $p_b > p_a$.\\

\begin{figure}[!hptb]
\begin{center}
\includegraphics[width=\columnwidth]{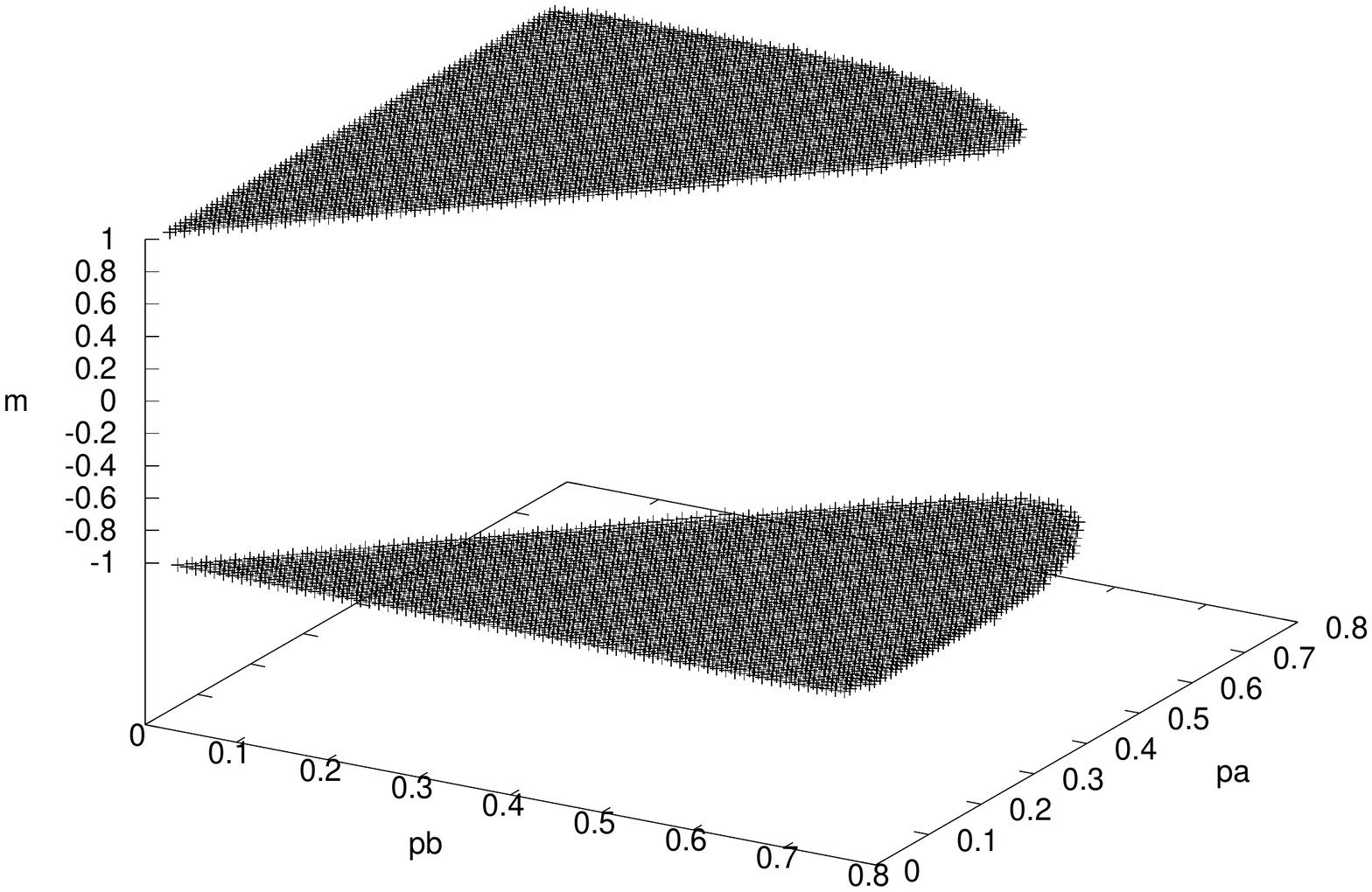}
\includegraphics[width=\columnwidth]{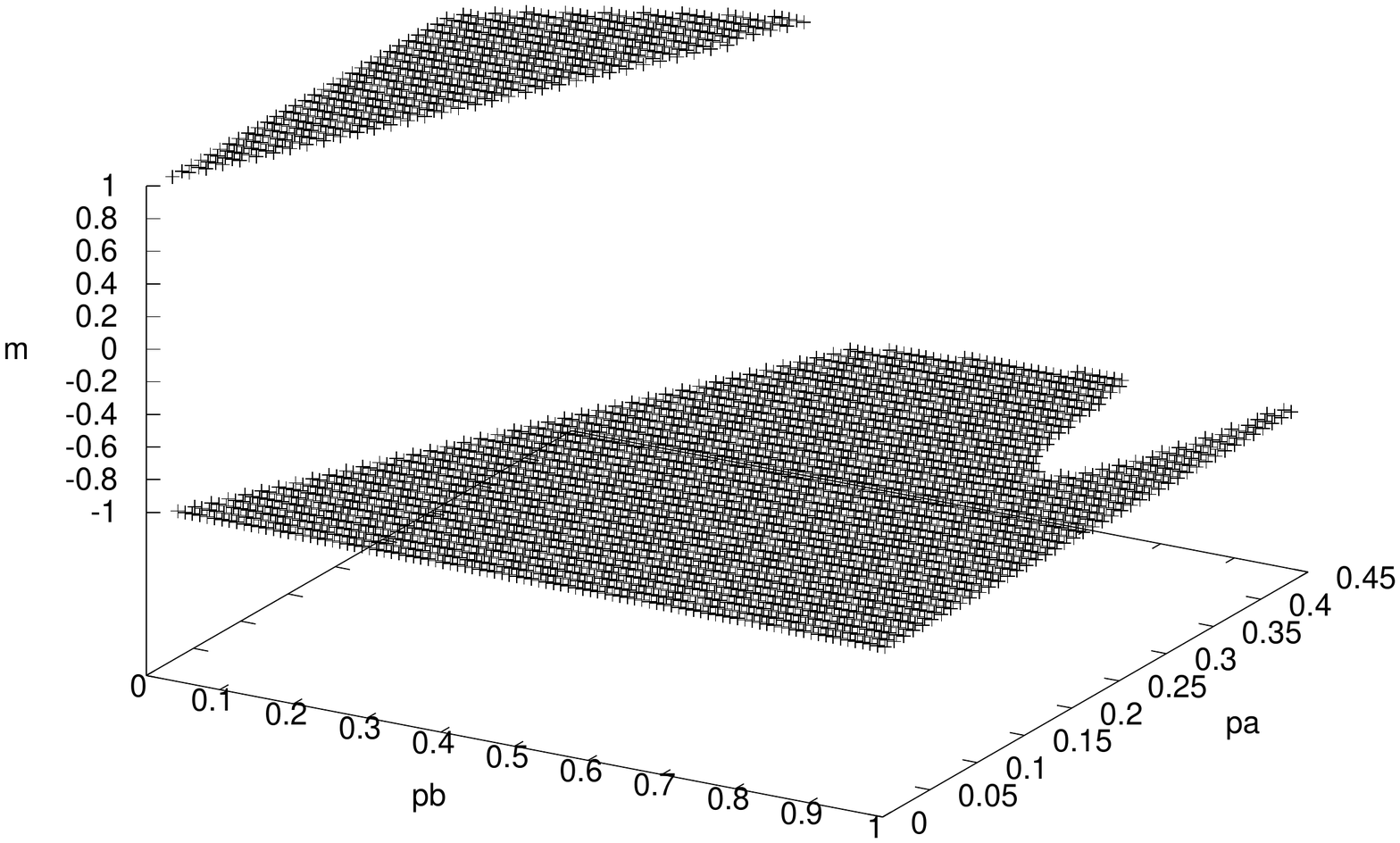}
\caption{The magnetization $m$ vs the position on the plane ($p_a,p_b$) for selected set of initial conditions: $\rho _0=0.5$, $m_0=0.0$, and $\beta _0=0.0$ (upper plot), $\rho _0=0.5$, $m_0=0.8$, and $\beta _0=-0.6$ (lower plot). For clarity, the plots are limited to the areas where $|m|>0.99$.}
\label{f5}
\end{center}
\end{figure}

In Fig. (\ref{f5}) two examples are shown of the areas where one of spin orientations is absent. For clarity of the picture, the surface plots are limited to the cases where
$|m|>0.99$. The upper plot is made for the symmetric system, the same as in the upper plot in Fig. (\ref{f4}). The lower plot in Fig. (\ref{f5}) is for the same data as the middle plot in Fig. (\ref{f4}). \\

\begin{figure}[!hptb]
\begin{center}
\includegraphics[width=\columnwidth]{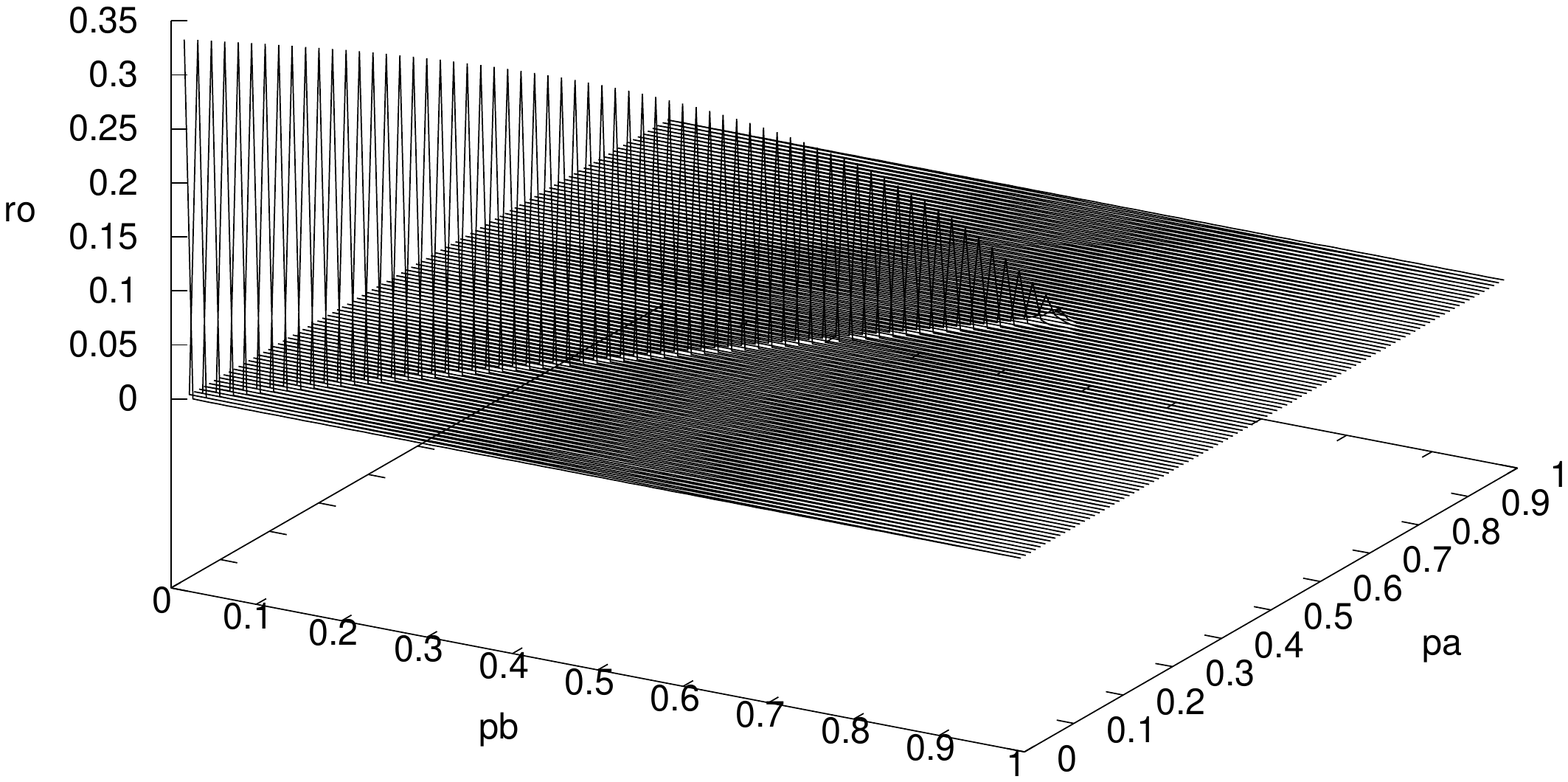}
\caption{The density of active links $\rho$ vs the position on the plane ($p_a,p_b$) for selected set of initial conditions: $\rho _0=0.5$, $m_0=0.0$, and $\beta _0=0.0$. The result is $\rho=0$ except at the line $p_b=p_a<2/3$. }
\label{f6}
\end{center}
\end{figure}

In Fig. (\ref{f6}) we show an exemplary dependence of the stationary density $\rho$ of active links against the positions in the plane ($p_b,p_a$). It is only for $p_a=p_b$, where $\rho$ is different from zero. This result is observed for all initial states ($\rho _0, m_0, \beta _0$).\\

\section{Discussion}
\label{S5}

When we treat the values of the probabilities $p_a$, $p_b$ as strategies of actors at nodes $a$ and $b$, we see that the community $a$ will not extinct as long as $p_a>p_b$.  Conversely, the community $b$ will not extinct as long as $p_a<p_b$. Once both communities are engaged in the game and both apply these conditions, both end at the state where $p_a=p_b=1$, i.e. they separate immediately, preserving their initial numbers $N_a,N_b$. Another solution is when one community copies the rewiring probability $p$ of another community. Then the results described in Figs. (\ref{f1},\ref{f2},\ref{f3}) remain valid.\\

We note that even in well-organized communities such collective decisions are rarely conscious. As we discussed in Section \ref{S1}, the patterns of separated communities emerge as unintended results of individual decisions. Following the classics, we could distinguish between 'community in itself' and 'community for itself', to conclude that the latter
is a premature reification \cite{oxf}. Actually the probabilities $p_a$, $p_b$ can be treated as strategies only within the evolutionary game theory \cite{egt}, and not as a conscious decision which optimizes an expected outcome.\\

For the modeling within the coevolving voter model, our results give two insights. First, following \cite{toru}, we demonstrate that for $p_a=p_b$ the final outcome of the time evolution of the system depends on its initial state. On the one hand, this result is interesting as an example of a collective memory in a social network, which remains out of sight of individual actors. On the other hand, the effect makes any systematic search of the model more difficult. It is tempting, for example, to determine the initial conditions $\rho_0, m_0, \beta_0$ which assure the transition to the frozen state $\rho=0$ continuous or discontinuous. Obviously, such research should be done for the whole range of the rewiring probability $p_a=p_b$. This task, far from being completed, is only mentioned here.\\

Our second insight is related to the case when $p_a \ne p_b$. Paradoxically, in this case the result is more clear: $\rho$ is equal to zero in all stationary states, unless a non-generic condition $p_a=p_b$ is fulfilled. For social interpretation, this result is disappointing, as it indicates that the coexistence of communities in mutual contact is not possible. It is fair to quote David Landes here: "Where one group is strong enough to push another around and stands to gain by it, it will do so." \cite{dsl}. In game theory, a similar frustration has been raised by the famous Prisoner's Dilemma: a proof that in given circumstances, cooperation is not reasonable for an individual \cite{strf}. We hope that the coevolving voter model will find further and more constructive generalizations.

%% ===========================================================================
\section*{Acknowledgements}
One of authors (K.K.) is grateful to Janusz Ho{\l}yst for hospitality and discussions, which initialized the calculations. This work was partly supported by the Faculty of Physics and Applied Computer Science (11.11.220.01/2) and by the Faculty of Humanities (11.11.430.158) AGH UST statutory tasks within subsidy of Ministry of Science and Higher Education.
%% ===========================================================================

\end{document}